\documentclass[a4paper,11pt]{article}
\usepackage[english]{babel}
\usepackage{float}
\usepackage[dvips]{graphicx}
\usepackage{amssymb}
\usepackage{amsmath}
\usepackage{cite}
\usepackage{amsthm}
\usepackage{amsopn}
\usepackage{multirow}
\usepackage{xcolor}
\usepackage[justification=centering]{caption}
\usepackage{latexsym}
\usepackage{epsfig,empheq}
\usepackage[mathscr]{eucal}
\usepackage{amsfonts}
\usepackage{amscd}
\usepackage{amsmath,amssymb}
\usepackage{amssymb}
\usepackage{colordvi}
\usepackage{graphicx}
\usepackage{fancyhdr} 
\usepackage{slashed}
\usepackage{tabularx}
\usepackage{longtable}
\usepackage{array}
\usepackage{hyperref}
\usepackage[autostyle]{csquotes} 
\usepackage{color}
\usepackage{ulem}

\newcommand{\beq}{\begin{eqnarray}}
\newcommand{\eeq}{\end{eqnarray}}

\newcommand{\GeV}{{\rm\ GeV}}

\newcommand{\inab}{\,{\rm ab}^{-1}}

\newlength{\myVSpace}
\begin{document}
\bibliographystyle{unsrt}
\begin{titlepage}
\begin{center}	
\vspace*{15mm}
\vspace{1cm}
{\Large \bf  Searching for lepton flavor violating interactions at  future electron-positron colliders }	\\			
\vspace{1cm}				
{\bf S. M. Etesami, R. Jafari, M. Mohammadi Najafabadi, S. Tizchang}\\
\vspace*{0.5cm}	
{\small  School of Particles and Accelerators, Institute for Research in Fundamental Sciences (IPM) P.O. Box 19395-5531, Tehran, Iran } \\
\vspace*{.2cm}
\end{center}
		
\begin{abstract}\label{abstract}
Lepton flavor violating interactions are absent in the standard model 
but are expected in various beyond standard models.
In this work, the potential of the future circular electron-positron collider to probe
the four fermion lepton flavor couplings via the  $e^{+}e^{-}\rightarrow e^{\pm}\tau^{\mp}$ process
is revisited by means of an effective field theory approach. 
We provide constraints at $95\%$CL on the dimension-six Wilson coefficients including
major sources of background processes and considering realistic detector effects at
four expected operation energies $\sqrt{s}=157.5$, $162.5$, $240$ and $365\GeV$ according to their 
corresponding integrated luminosities. 
We demonstrate that  statistical combination of the results from four center-of-mass energies, 
improve the sensitivity to the LFV couplings significantly.
We compare the results with the prospects from Belle II with $50 \inab$
and other studies at electron-positron colliders.
\end{abstract}
\vspace*{3mm}

\end{titlepage}
\section{Introduction}

In the standard model (SM) with massless neutrinos, processes with lepton flavor violating (LFV) interactions are 
forbidden \cite{Glashow:1961tr}. However, experimental observations of neutrino oscillations show that neutrinos are massive 
and mix with each other which leads to violation of lepton flavor conservation \cite{Weinberg:1967tq}.
LFV enters into the charged lepton sector from the neutrino sector via radiative corrections which
are extremely suppressed because of the smallness of the ratio of neutrino mass to the W boson mass \cite{r1,r2}.
The predicted branching fraction for example for $\tau^{-} \rightarrow \ell^{+} \ell^{-} \ell'^{-}$ decays are approximately
$\lesssim 10^{-54}$, where $\ell = e, \mu$ \cite{r2}. 
However, an increase of several orders of magnitude is predicted in some extensions of the SM,
such as supersymmetric SM \cite{r3,r4,r5}, resulting in branching fractions observable at experiments.
Therefore,  any  observation of LFV in the charged lepton sector
would be an obvious hint to the presence of physics 
beyond  the SM,  and can be an indirect way to search for beyond the SM scenarios.

So far, no LFV interactions among the charged leptons have been observed
and there are several strong constraints from various experiments. 
For instance,  $\ell ^{-} \rightarrow e^{-}e^{+}e^{-}$, $\ell \rightarrow e\gamma$ with $\ell = \tau,\mu$,  
rare decays of mesons, $Z$ boson decays, decays of Higgs boson and heavy resonances 
have been used to probe LFV in different experiments \cite{{Aad:2016wce},{Aaij:2014azz},{Aad:2016blu},{Aaij:2015qmj}, {rr1},{rr2},{rr3},{rr4},{rr5},{rr6}}.
The most stringent  bounds  at $90\%$ CL on the LFV decays of the
$\tau$ leptons into $3e$ were measured by BaBar and Belle \cite{r6,r7}:
\begin{eqnarray}
\mathcal{B}(\tau^{-}\rightarrow e^{-}e^{+}e^{-}) \leq 2.9 \times 10^{-8}~(\text{BaBar})~,~\mathcal{B}(\tau^{-}\rightarrow e^{-}e^{+}e^{-}) \leq 2.7 \times 10^{-8}~(\text{Belle}),
\end{eqnarray} 
The Belle II future prospects for upper limit on $\mathcal{B}(\tau^{-}\rightarrow e^{-}e^{+}e^{-})$
at $90\%$ CL assuming the integrated luminosity of 50 ab$^{-1}$ is $\lesssim 10^{-10}$ \cite{r8}.

The proposed future lepton colliders such as
the International Linear Collider (ILC)~\cite{Aihara:2019gcq,Baer:2013cma,Behnke:2013xla}, the Compact Linear Collider (CLIC)~\cite{Charles:2018vfv,Roloff:2018dqu,Aicheler:2019dhf}, 
Circular Electron-Positron Collider (CPEC) \cite{CEPC-SPPCStudyGroup:2015csa,CEPC-SPPCStudyGroup:2015esa} 
and  Future Circular Collider with electron-positron collisions  
(FCC-ee) with highest-luminosity \cite{Abada:2019zxq}  
are expected to provide an extraordinary place to perform flavor physics studies.
There are a variety of theories that give rise to LFV. 
For instance,  additional fermions present in the type III seesaw model or in the
low-scale seesaw models give rise to large LFV effects
\cite{Esteves:2009vg, Cai:2017mow, Goswami:2018jar,{Das:2014jxa},{Das:2015toa},{Das:2016hof},{Das:2017nvm},{Das:2018usr},{Das:2017gke}}.
 The LFV through $Z$, Higgs boson, and other new degrees of freedom have been studied 
 in Refs.\cite{{Delepine:2001di},{Illana:2000ic},{FloresTlalpa:2001sp},{Perez:2003ad},{Cirigliano:2021img},{Buras:2021btx},{Crivellin:2020ebi}}, 
and Higgs and scalar LFV decays have been presented in  Refs.\cite{{Dev:2017ftk},{Dev:2019ugu},{Goudelis:2011un},{Vicente:2019ykr},{Crivellin:2013wna}}. 
If the new degrees of freedom contributing to LFV are heavy comparing to the energy accessible at colliders
then the LFV couplings could be reasonably parameterized via the effective contact interactions. 
Experiments can perfectly search for LFV in a model-independent approach, without any theoretical input. 
Effective field theories allow for a model-independent interpretation of the experimental results. 
However, in a true bottom-up approach, all relevant operators have to be considered 
since no symmetry or model consideration is present to suppress some operators with respect to other operators.

LFV $\bar{e}e\bar{e}\tau$ contact interactions have been already studied at future high energy 
lepton colliders through $e^{+}e^{-} \rightarrow e^{\pm} \tau^{\mp}$ in Refs.\cite{lfv1,lfv2}.
In Ref.\cite{lfv2}, the LFV contact operators probed via $e^{+}e^{-} \rightarrow e^{\pm} \tau^{\mp}$
process at $\sqrt{s} = 250, 500, 1000, 3000$ GeV considering two main background sources, $\tau^{+}\tau^{-}$
and $e\tau \nu_{e}\nu_{\tau}$. Similar process examined  at $\sqrt{s} = 250, 500, 1000$ GeV  in Ref.\cite{lfv2}
where the effects of polarization of the electron and positron beam have been investigated. 
The detector response has been simulated using {\tt Delphes} package  \cite{deFavereau:2013fsa}
and $e^{\pm}\tau^{\mp} \nu_{e}\nu_{\tau}$ has been considered as the main source of background.

The purpose of this paper is to investigate the LFV contact interactions of
 $e\bar{e}e\bar{\tau}$ at a future lepton collider. In particular, the focus is on a model independent search 
 for various types of four-lepton LFV
interactions leading to the production of $ e^{\pm} \tau^{\mp}$ at the future circular electron-positron collider.
The search is performed at the proposed energies and integrated luminosity benchmarks of the
FCC-ee. Particularly, the analysis is carried out at the center of mass energies 
of $157.5$, $162.5$, $240$ and $365\GeV$ with the corresponding integrated luminosities of 
$5, 5, 5,$ and $1.5$ ab$^{-1}$, respectively. A realistic detector response is taken into account using
{\tt Delphes} package. The main sources of background processes are
$\tau^{+}\tau^{-}$,  $e^{\pm}\tau^{\mp} \nu_{e}\nu_{\tau}$, $\ell^\pm \ell^ \mp \ell'^\pm \ell'^\mp$,
$\ell^\pm \ell^\mp jj, \ell\nu jj, (\ell=e, \mu,\tau)$, and $jj, (j = \text{jet})$.
Finally, a statistical combination is performed over the results obtained at the four center-of-mass energies.

The layout of this paper is as follows. In section \ref{sec:II} the effective operators describing LFV 
are briefly presented.
Section \ref{sec:III} is dedicated to present the
simulation details and  analysis strategy.
In section \ref{sec:IV}, the constraints on LFV couplings and the results of statistical combinations
are given. Finally, section \ref{sec:V} concludes the paper.
		 
\section{The LFV effective Lagrangian} \label{sec:II}
 
For studying the $\bar{e}e\bar{e}\tau$ LFV couplings, it is customary in the literature to consider 
four fermi contact interactions which provide the opportunity to
characterize the new physics effects in a 
model-independent framework. 
In general, there is a set of six chirality conserving scalar and vector form four-fermi operators 
with  $\Delta L=1$ where $\Delta L$ represents the difference between initial and final state lepton numbers \cite{Kuno:1999jp}.
The operators are classified into two types: the scalar type $(S)$ and vector type ($V)$ interactions. 
In addition, there are LFV operators containing dipole structures which are tightly constrained by radiative LFV 
decays \cite{Tanabashi:2018oca}, therefore, they are not considered in this work. 
The effective Lagrangian and the relevant set of operators leading to $e^{\pm} \tau^{\mp}$ production by either a scalar or vector are given by \cite{Kuno:1999jp}: 
	\begin{align}
	\label{eq:coupling}
	\mathcal{L}_{\mathrm{eff}} & \supset \sum_{\alpha,\beta}\sum_{ij}\dfrac{c_{\alpha\beta}^{ij}}{\Lambda^2}\, \mathcal{O}_{\alpha\beta}^{ij}\,,
	\end{align}
	\begin{eqnarray}
		\label{eq:coupling1}
	\mathcal{O}^{S,ij}_{RL}&=& \left(\overline{\ell}_{jL}{\ell}_{iR}\right) \left(\overline{\ell}_{jL}{\ell}_{jR} \right),\,\,\,\,\,\,\,\,\,\,\,\,\,\,\,\,\,\,\,\,\,\,\,\, \mathcal{O}^{S,ij}_{LR}= 
	\left(\overline{\ell}_{iR}{\ell}_{jL} \right) \left(\overline{\ell}_{jR}{\ell}_{jL}\right),\nonumber\\
	\mathcal{O}^{V,ij}_{RR}&=&\left(\overline{\ell}_{iR}\gamma^\mu {\ell_{jR}} \right)
	\left(\overline{{\ell}}_{jR}\gamma_\mu {\ell_{jR}} \right) ,\,\,\,\,\,\,\,\,\,\,
	\mathcal{O}^{V,ij}_{LL}=\left(\overline{\ell}_{iL}\gamma^\mu \ell_{jL} \right)
	\left(\overline{\ell}_{V,jL}\gamma_\mu \ell_{jL} \right),\nonumber \\
	\mathcal{O}^{V,ij}_{LR}&=&\left(\overline{\ell}_{iL}\gamma^\mu {\ell}_{jL} \right),
	\left(\overline{\ell}_{jR}\gamma_\mu {\ell}_{jR} \right),\,\,\,\,\,\,\,\,\,\,
	\mathcal{O}^{V,ij}_{RL}= \left(\overline{\ell}_{iR}\gamma^\mu {\ell}_{jR} \right)
	\left(\overline{\ell}_{iL}\gamma_\mu {\ell}_{iL} \right),	   	
	\end{eqnarray}
where $\mathcal{O}_{\alpha\beta}^{ij}$ are the four fermion leptonic operators, $\Lambda$ is the new physics energy scale, and 
$c_{\alpha\beta}^{ij}$s indicate the effective Wilson coupling  between leptons of flavor $i$ and $j$
and $\alpha\beta$  Lorentz structures. The operators are invariant under the SM 
gauge symmetry $\rm SU(3)\times SU(2)\times U(1)$. It is found that flavor violation among first and 
second generations of leptons is tightly constrained by experimental constraints arising from  
muon decay into three electron $\mu \rightarrow e e e$ at SINDRUM experiment\cite{Bellgardt:1987du}, 
the muon transition to $e \gamma$ \cite{TheMEG:2016wtm}, and $\mu-e$ conversion \cite{Bertl:2006up}. 
However, constraints on flavor violations between electron and $\tau$ and, muon and $\tau$ are much looser. 
As a consequence, we restrict our study to $\bar{e}e\bar{e}\tau$ couplings  using $e^{-}e^{+} \rightarrow e^{\pm} \tau^{\mp}$ process.

In addition to the $\bar{e}e\bar{e}\tau$ four fermi contact interactions, 
contact interactions among leptons and quarks (like $\bar{e}e\bar{q}q'$), and electrons and Higgs-$Z$ ($eeHZ$)
are of favourite topics which have been probed in several papers such as Refs. \cite{c1,c2,c3,c4,Crivellin:2013hpa}.

In order to see the dependence of the production rate to the center of mass energy $\sqrt{s}$
and to find a feeling about the sensitivity to different types of operators, the expression for the cross section
of $e^{-}e^{+} \rightarrow e^{\pm} \tau^{\mp}$ process is presented.  
The theoretical cross section of $\sigma(e^{-}e^{+} \rightarrow e^{+} \tau^{-})+\sigma(e^{-}e^{+} \rightarrow e^{-} \tau^{+})$ 
in the presence of all couplings has the following form \cite{Ferreira:2006dg}:
\begin{eqnarray}
\sigma \left( s \right) & = &
\frac{s}{96\pi \Lambda^4}
\Big\{( |c^S_{LR}|^2 + |c^S_{RL}|^2)+ 16 (|c^V_{LL}|^2 + |c^V_{RR}|^2 +  |c^V_{LR}|^2 + |c^V_{RL}|^2) \Big\} .~~~
\label{eq:sigma}
\end{eqnarray}
where in finding the above expression, the lepton masses are set to zero considering the center-of-mass energy scale.
As seen, production rate of the four-fermion interactions grows linearly with the squared center-of-mass energy $s$, 
and diverge when $s \rightarrow \infty$. However, one should note that we are working in a non-renormalizable formalism
and  these operators provide an acceptable description of physics at high energy up to an energy scale $\Lambda$.
Another interesting point which is worth mentioning is that the vector type operators contribute to LFV production
of $e\tau$ with a factor of 16 larger than the scalar type operators. Therefore, better sensitivity is expected 
to vector type with respect to scalar type operators.

In the next section, we describe the simulation method and details of the analysis 
to search for the LFV operators at four energy benchmarks of the FCC-ee.

\section{ Simulation details and analysis strategy}\label{sec:III}

The main goal of this work is to estimate the potential of the  proposed  FCC-ee collider to probe the 
effective LFV operators introduced in Eq.\ref{eq:coupling} using $e^{-}e^{+} \rightarrow e^{\pm} \tau^{\mp}$ process. 
This section is dedicated to present event generation, simulation of detector effects 
and analysis method to find the LFV sensitivity for four energy benchmarks.
In particular, the search is separately  performed at  center-of-mass energies 
${365}$, ${240}$, ${162.5}$ and ${157.5}$ GeV with their expected 
integrated luminosities of $1.5$, $5$, $5$ and $5 \inab$ at FCC-ee collider, respectively \cite{Abada:2019zxq}. 
The signal process consists of an electron (or a positron) and a $\bar{\tau}$ lepton (or a $\tau$ lepton) which decays hadronically.  
The main background sources which are taken into account in this study are:
\begin{eqnarray*}
	\text{(I)}   	&&   e^- e^+\rightarrow e^\pm\tau^\mp\nu \bar{\nu},\\
	\text{(II)}       &&  e^- e^+\rightarrow \tau^{+}\tau^{-}, 	\\
	\text{(III)}	&&  e^- e^+\rightarrow \ell^\pm \ell^ \mp \ell'^\pm \ell'^\mp\, (\ell, \ell'=e,\mu,\tau),\\
	\text{(IV)}	&&  e^- e^+\rightarrow \ell^\pm \ell^\mp jj\ (\ell=e,\mu,\tau), \\
	\text{(V)} 	&& e^- e^+\rightarrow \ell^\pm\nu jj (\ell=e,\mu,\tau),\\
		\text{(VI)}	&&  e^- e^+\rightarrow jj.
\end{eqnarray*} 
The second background, $\tau^{+}\tau^{-}$,  is in particular contributing to 
the background composition when one of the $\tau$ leptons decay to an electron and the electron 
is reconstructed in the final state. 
The third item in the list of backgrounds, $\ell^\pm \ell^ \mp \ell'^\pm \ell'^\mp$, is considered as there is a possibility that
 two of the leptons are scattered to regions of pseudorapidity where detector area is blind or
 two isolated charged leptons are not well reconstructed. 
Since jets can be misidentified as electrons and also as
$\tau$ leptons, the fourth to sixth items
must be included in the list of backgrounds to obtain a more realistic assessment of the results.

The effective  Lagrangian introduced in Eq. \ref{eq:coupling} 
is implemented in the {\tt FeynRule} program \cite{Alloul:2013bka} and then the Universal FeynRules Output (UFO) model \cite{ufo}
is inserted to {\tt MadGraph5\_aMC@NLO 2.6.6}. The events of signal and backgrounds 
are generated at leading order with {\tt MadGraph5\_aMC@NLO  2.6.6} \cite{Alwall:2011uj,Alwall:2014bza,Alwall:2014hca} 
including ISR effect. The ISR effects are considered using the {\tt MGISR} plugin in {\tt MadGraph5\_aMC@NLO  2.6.6} \cite{{Chen:2017ipx},{Li:2018qnh}}.  
The generated events are passed through {\tt PYTHIA 8} \cite{Sjostrand:2007gs,Sjostrand:2014zea} 
for showering, hadronization, and decay of unstable particles.
The detector effects are simulated via {\tt Delphes 3.4.2}~\cite{deFavereau:2013fsa} 
according to an ILD-like detector \cite{Behnke:2019sdv}.
For electrons with $p_{T} > 10$ GeV and $|\eta| \leq 2.5$, the identification efficiency in the ILD card
is $95\%$. The electron energy resolution is: $\frac{\Delta E}{E} = \frac{0.15}{\sqrt{E}} + 0.01$.
To calculate the cross section of signal and backgrounds and simulate the events,
the SM relevant input values are set as follows: $M_{Z} = 91.188$ GeV,
mass of $\tau$ lepton $m_{\tau} = 1.777$ GeV, $G_{F} = 1.166\times 10^{-5}$ GeV$^{-2}$,
$\alpha_{e} = 1/127.9$, and $\alpha_{s} = 0.118$.

The  $\tau$ lepton lifetime is  $\sim 290$ femto second, corresponding to $c\tau \sim 87 \mu m$.
The actual decay  length is obtained by multiplying $c\tau$  with $\beta\gamma$ therefore
a $\tau$ lepton with $E=40$ GeV travels around $2$ mm through the detector then decays.  
 In  around two  thirds of decays, $\tau$'s decay  hadronically, 
 typically to one or three charged mesons (mostly $\pi^{-}\pi^{+}$), 
 often associated with neutral pions, and a $\tau$ neutrino.
The $\tau$ tagging efficiency in the ILD simulation card is $40\%$ 
and the $\tau$ misidentification rate is assumed to be equal $0.1\%$ \cite{potter}. 

 Six different signal samples  corresponding to the six operators presented in Eq.\ref{eq:coupling1} 
 are generated. In order to generate signal events, we set related effective Wilson coefficient $c_{ij} = 0.1$, with $i=j=L,R$, and $\Lambda=1 $ TeV
 and require pre-selection cuts as $p_T^{\ell}=10\GeV$, pseudorapidity of leptons $|\eta^{\ell}|<3.0$ and angular separation between 
 two leptons as $\Delta R = \sqrt{(\Delta\eta)^{2} + (\Delta\phi)^{2}} > 0.3$. 
 For generating the background samples, we require the  addition of angular separation between 
 leptons and jets, and two jets in the final state to be  $\Delta R_{\ell,j}>0.3$ and the $p_T=10\GeV$ and $|\eta|<3.0$ 
 are applied on jets in the final state at generator level.
 
The cross-section and corresponding uncertainties with ISR effect of  backgrounds 
as well as signals are given in Table \ref{Table:xsec} for all center-of-mass energies $157.5$, $162.5$, $240$ and $365$ GeV. 

\begin{table*}[]
\begin{center}
\tiny
\begin{tabular}{c|cc|cccccc}
\hline \hline
$\sqrt s$ [GeV]    & $c^V_{LR}=0.1$ & $c^S_{LR}=0.1$ & $e \tau\nu\bar{\nu}$ & $\tau\bar\tau $ &  $\ell \bar{\ell }\ell'\bar{\ell' } $ & $\ell \bar{\ell }jj$ & $\ell \nu jj$ & $jj$ \\ \hline  \hline
157.5 &   $4.72\pm0.007$  & $0.29\pm0.0004$ &  $22.33\pm0.07$  &  $11076.5\pm3.4$  & $39.86\pm0.08$ & $80.95\pm0.2$ & $272.9\pm0.4$  & $32032\pm8.1$\\\hline
162.5 &   $5.02\pm0.007$  & $0.31\pm0.0004$ &  $102.12\pm0.3$ &  $10275.8\pm2.9$  & $42.23\pm0.08$ & $83.06\pm0.3$ & $1198.05\pm0.8$ & $29133\pm6.2$ \\	\hline 
240   &   $10.98\pm0.04$ & $0.69\pm0.0008$ &  $415.63\pm0.6$ &  $4196.8\pm1.2$   & $86.24\pm0.2$ & $217.8\pm0.5$ & $4552.7\pm1.3$ & $10481\pm3.5$\\	\hline 
365   &   $25.26\pm0.07$ & $1.57\pm0.002$ &  $327.59\pm0.5$ &  $1803.6\pm0.6$  & $85.05\pm0.1$ & $195.13\pm0.3$ & $3247.02\pm1.1$ & $4306\pm1.2$ \\			 \hline \hline
\end{tabular}
	\end{center}
	\caption{\small The cross sections of signal $e^- e^+ \rightarrow e^{\pm} \tau^{\mp}$ and main background processes with
	their corresponding uncertainties are presented.
	 The cross section of two signal scenarios are given assuming  $c^V_{LR}=0.1$, $c^S_{LR}=0.1$, and  $\Lambda = 1$ TeV.  
	 The cross sections are in unit of fb and are after including the ISR effects.}
	\label{Table:xsec}
\end{table*}


The  signal events are selected  according to the following requirements.
We  ask for exactly  two leptons with opposite charge, one isolated electron  and one $\tau$-tagged  lepton.
It is required that  $p_T^{e}>10 \GeV$, $p_T^{\tau}>20 \GeV$  and pseudorapidity of both leptons must satisfy $|\eta^{\ell}|<2.5$. and $\Delta R_{e,\tau} > 0.5$.
In order to  make sure the electron candidate is well isolated, it is required $\text{RelIso} < 0.15$, where 
$\text{RelIso}$ is defined as the ratio of the sum of $p_{\rm T}$ of charged particle tracks inside a cone of size
$0.5$ around the electron track to $p_{\rm T}$ of the electron. 
$\tau$ leptons which decay via hadronic  modes are considered.
A $\tau$ lepton in hadronic  decay mode produces a jet containing a few neutral and charged hadrons.
 Therefore, considering the $\tau$ tagging efficiency, a jet is considered potentially as a $\tau$ candidate if
 a generated $\tau$ exists within a  distance
 $\Delta R = \sqrt{(\eta_{\rm jet} - \eta_{\tau})^{2} + (\phi_{\rm jet} - \phi_{\tau})^{2}} = 0.3$
 from the jet axis. 
 
 To suppress the contributions from events with $e^{-}e^{+}$ and $\mu^{-}\mu^{+}$ in the final state,
events that contain two electron or two muon candidates are rejected. 
This reduces background events from $\ell^\pm \ell^ \mp \ell'^\pm \ell'^\mp$, 
 $\ell^\pm \ell^\mp jj$, and  $\tau^{+}\tau^{-}$ with $\tau \rightarrow e \nu_{e} \nu_{\tau}$
 or $\tau \rightarrow \mu \nu_{\mu} \nu_{\tau}$.
 
\begin{figure*}[htb]
\begin{center}
 {\includegraphics[ scale=0.4]{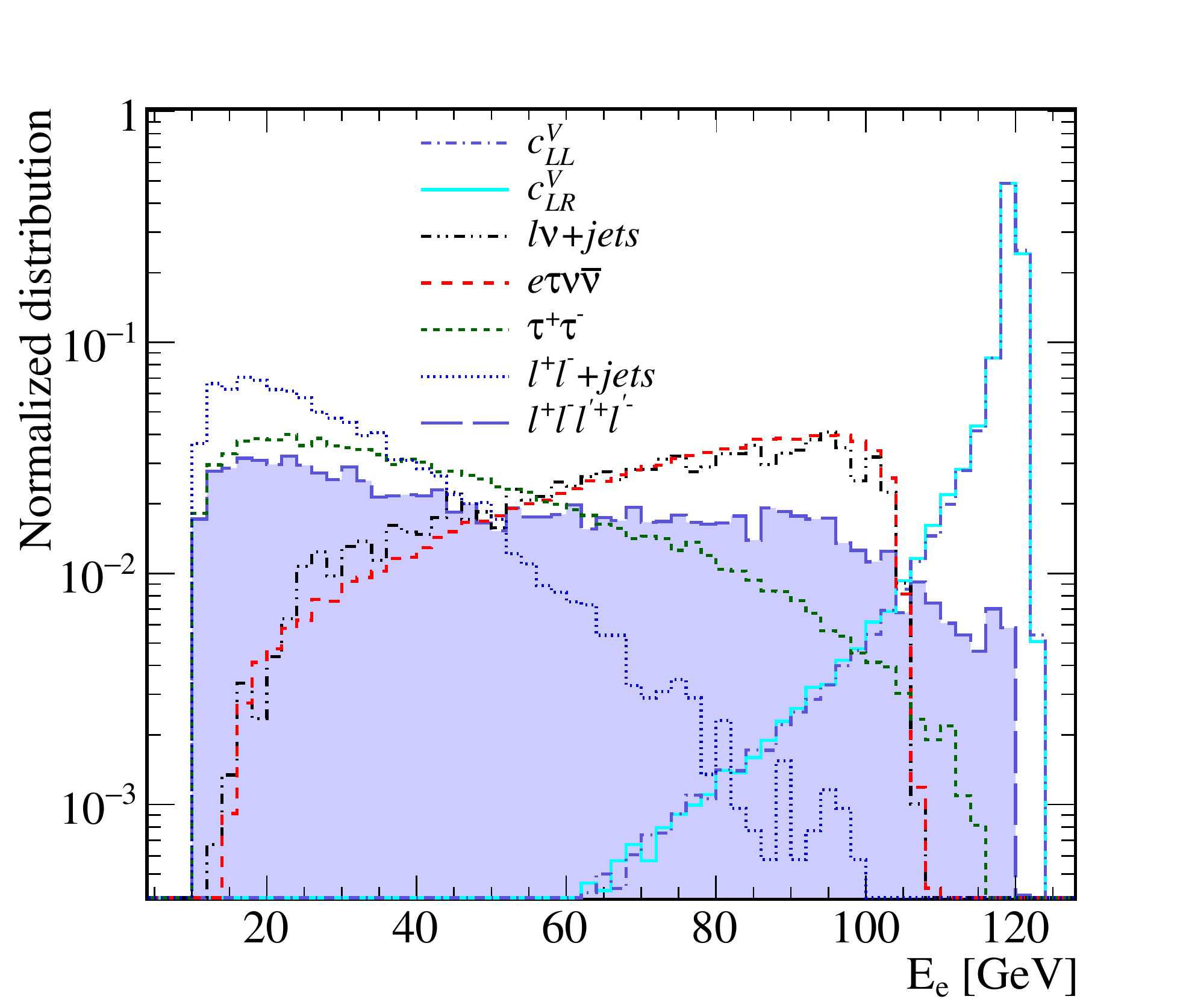}}
 {\includegraphics[ scale=0.4]{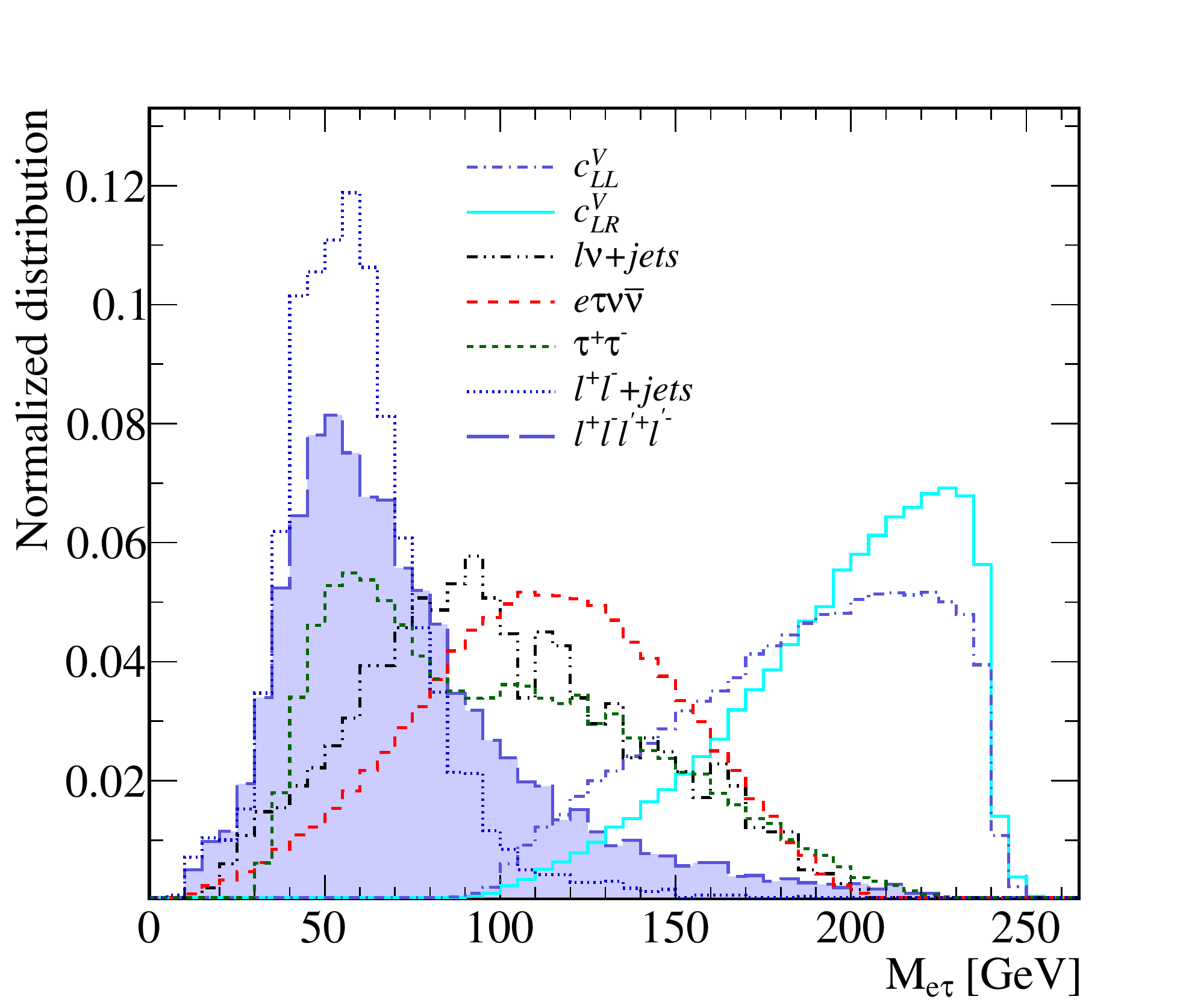}}
 \caption{{\small  Normalized distributions of electron energy(left), 
 and invariant mass of two final state charged leptons $M_{e\tau}$ (right) 
 are presented  for  signal benchmarks  $c^V_{LL}=0.1$ and $c^V_{RL}=0.1$, 
 at $\sqrt{s} = $ 240 GeV.}} \label{distribution}
  \end{center}
  \end{figure*}

 To enhance the sensitivity, we apply additional cuts on the  energy of final state electron 
 and the invariant mass of the two leptons $M_{e\tau}$. 
 Figure \ref{distribution} displays the distributions of the electron energy in the final state (left plot)
 and invariant mass of final state, i.e. $M_{e\tau}$ (right plot) at a center-of-mass energy of 240 GeV.
 Obviously,  both distributions show  significant difference between 
 the shape of signal and background processes.
 The $M_{e\tau}$ distributions for $\ell^\pm \ell^\mp jj$, $\tau^+ \tau^-$ and $\ell^\pm \ell^ \mp \ell'^\pm \ell'^\mp$
 background processes peak at lower values with respect to the signal and even
 the other backgrounds. This is expected as the center-of-mass energy is shared 
 among at least four particles in $\ell^\pm \ell^\mp jj$ and $\ell^\pm \ell^ \mp \ell'^\pm \ell'^\mp$ processes.
 Therefore, cutting on  $M_{e\tau}$  suppresses effectively these reducible backgrounds.
 It is notable that the $M_{e\tau}$ distribution has different behaviours 
 for $c^{V}_{LL}$ and $c^{V}_{LR}$ which arises from the fact that
 for $LL$ coupling $d\sigma/d\cos\theta \propto (1+\cos\theta)^{2}$
 while for $LR$ coupling $d\sigma/d\cos\theta \propto (1-\cos\theta)^{2}$.
 For the signal events, the energy distribution of the electron is expected to spread around $\sqrt{s}/2$.
 As a result, it is used to further reduce the contributions of the background processes.

 The cut values on $E_{e}$ and $M_{e\tau}$ are optimized such that
 the best limit on the LFV coefficients are obtained. In order to optimize the cuts
  on the invariant mass and energy of the outgoing electron, 
   the upper bound on the signal cross section for 
  different cut values on $M_{e\tau}$ and $E_{e}$ is obtained.  
  The values which gives lowest upper bound on signal cross section are chosen as the optimum values. 
 The  cuts on invariant mass of final state $M_{e\tau}$ are found to be greater than 
 65$\GeV$, 100 and 150$\GeV$ for $\sqrt{s}=157.5 - 162.5$, $240$ and $365\GeV$, respectively. 
 The optimized lower cuts on energy of electron are obtained to be  
 $78.6$, $81.0$, $119.7$ and $182.0 \GeV$  for $\sqrt{s}=157.5$, $162.5$, 240 and $365$ GeV, respectively.
 Table \ref{Table:eff} presents the efficiencies of signal with $c^V_{LR}=0.1$, $c^S_{LR}=0.1$ and 
the main SM backgrounds after preselection cuts, $M_{e\tau}$, and $E_{e}$ 
with $\sqrt{s}=157.5$, $162.5$, 240 and $365\GeV$.
 After applying the cuts, the main background contributions arise from
 $\ell^\pm \ell^ \mp \ell'^\pm \ell'^\mp$,  $\tau^{+}\tau^{-}$ and $\ell^{\pm} \ell^{\mp} jj$ and the rest are remarkably suppressed.

\begin{table*}[]
	\begin{center}
		\scalebox{0.85}{
			\begin{tabular}{c |c c|c c c c c}
				\hline \hline
				\multirow{2}{*}{$\sqrt s = 157.5$ GeV} & \multicolumn{2}{c|}{Signal} & \multicolumn{4}{c}{~~~~~~ SM Backgrounds }\\ \cline{2-8}
				& $c^V_{LR}=0.1$ & $c^S_{LR}=0.1$ & $e \tau\nu\bar{\nu}$ &   $\tau\bar\tau $ &  $\ell \bar{\ell }\ell'\bar{\ell' } $ & $\ell \bar{\ell }jj $ & $\ell\nu jj$ \\ \hline 
				(I): Pre-selection cuts   &   $0.1746$  & $0.1698$ &  $0.099$  &  $0.045$  & $4.9\times 10^{-3}$ & $1.4\times 10^{-3}$ & $3.3\times 10^{-4}$ \\
				(II): $M_{e \tau}>65 $ GeV &   $0.1741$  & $0.1697$ &  $0.038$  &  $0.019$  & $2.2\times 10^{-3}$ & $1.8\times 10^{-4}$  & $7.5\times 10^{-5}$ \\		
				(III): $E_e > 78.6$ GeV &  $0.0984$  & $0.0831$ &$2.8\times 10^{-8}$ & $1.5\times 10^{-7}$  &   $6.02\times 10^{-6}$  &  $1.7\times 10^{-7}$ & $ 0.0$  \\		
				\hline  \hline				
				\multirow{2}{*}{$\sqrt s = 162.5$ GeV} & \multicolumn{2}{c|}{Signal} & \multicolumn{4}{c}{~~~~~~ SM Backgrounds }\\ \cline{2-8}
				& $c^V_{LR}=0.1$ & $c^S_{LR}=0.1$ & $e \tau\nu\bar{\nu}$ &   $\tau\bar\tau $ &  $\ell \bar{\ell }\ell'\bar{\ell' } $ & $\ell \bar{\ell }jj $  & $\ell\nu jj$\\ \hline 
				(I): Pre-selection cuts   &   $0.1727$  & $0.1711$ &  $0.106$  &  $0.048$  & $4.9\times 10^{-3}$ & $1.6\times 10^{-3}$ & $4.5\times 10^{-4}$  \\
				(II): $M_{e \tau}>65 $ GeV &   $0.1727$  & $0.1710$ &  $0.041$  &  $0.025$  & $2.4\times 10^{-3}$ & $2.1\times 10^{-4}$ & $1.0\times 10^{-4}$ \\		
				(III): $E_e > 81$ GeV &  $0.1122$  & $0.0949$ &$6\times 10^{-8}$ & $2.0\times 10^{-7}$  &   $3.61\times 10^{-6}$  &  $2.1\times 10^{-7}$ & $0.0$ \\			
				\hline  \hline				
				\multirow{2}{*}{$\sqrt s = 240$ GeV} & \multicolumn{2}{c|}{Signal} & \multicolumn{4}{c}{~~~~~~ SM Backgrounds }\\ \cline{2-8}
				& $c^V_{LR}=0.1$ & $c^S_{LR}=0.1$ & $e \tau\nu\bar{\nu}$ &   $\tau\bar\tau $ &  $\ell \bar{\ell }\ell'\bar{\ell' } $ & $\ell \bar{\ell }jj $  & $\ell\nu jj$\\ \hline 
				(I): Pre-selection cuts   &   $0.2156$  & $0.2137$ &  $0.131$  &  $0.037$  & $8.8\times10^{-3}$ & $6.2\times10^{-3}$ & $4.9\times10^{-4}$ \\
				(II): $M_{e \tau}>100 $ GeV &   $0.2150$  & $0.2134$ &  $0.084$  &  $0.017$  & $1.6\times10^{-3}$ & $2.4\times10^{-4}$  &  $2.0\times10^{-4}$\\						
				(III): $E_e > 119.7$ GeV &  $0.1072$  & $0.0989$ &$2.1\times 10^{-8}$ & $1.5\times10^{-7}$  &   $1.2\times 10^{-5}$  &  $2.4\times10^{-7}$ & $ 0.0$  \\		
				\hline  \hline
				\multirow{2}{*}{$\sqrt s = 365$ GeV}& \multicolumn{2}{c|}{Signal} & \multicolumn{4}{c}{~~~~~~ SM Backgrounds }\\ \cline{2-8}
				& $c^V_{LR}=0.1$ & $c^S_{LR}=0.1$ & $e \tau\nu\bar{\nu}$ &   $\tau\bar\tau $ &  $\ell \bar{\ell }\ell'\bar{\ell' } $ & $\ell \bar{\ell }jj $  & $\ell\nu jj$\\ \hline 	
				(I): Pre-selection cuts   &   $0.2093$  & $0.2097$ &  $0.133$  &  $0.066$  & $0.012$ & $6.0\times 10^{-3}$ & $5.0\times 10^{-4}$\\
				(II): $M_{e \tau}>150 $ GeV  &   $0.2053$  & $0.2051$ &  $0.093$  &  $0.041$  & $2.0\times 10^{-3}$ & $1.5\times 10^{-4}$ & $2.4\times 10^{-4}$ \\							
				(III): $E_e > 182$ GeV &  $0.0993$  & $0.0986$ &$2.6\times 10^{-8}$ & $3.2\times10^{-7}$  &   $2.6\times 10^{-5}$  &  $1.4\times 10^{-7}$ & $ 0.0$  \\			
				\hline  \hline
			\end{tabular}}
		\end{center}
		\caption{The efficiencies for signal with $c^V_{LR}=0.1$, $c^S_{LR}=0.1$ and 
					the SM backgrounds after selection cuts with $\sqrt{s}=157.5$, $162.5$, 240 and $365\GeV$ are given.}
		\label{Table:eff}
	\end{table*}

As previously indicated, jets could be misidentified as 
$\tau$ and electron therefore, processes with jets in the final state 
contribute to the background. 
The detectors proposed for the future lepton colliders are
expected to have a great performance better than the current modern 
multipurpose detectors such as ATLAS and CMS detectors. 
The jet fake $\tau$ probability 
 is  expected to be $0.1\%$ \cite{potter}. 
The rate of background containing jets varies with the center-of-mass energy and 
is assessed to be less than $5\%$
of the total
background contributions after all selection criteria. 

In the next section, we evaluate the potential sensitivities to LFV
operators at four energy benchmarks. 
In addition,  a statistical combination of four center-of-mass energies is presented.


\section{ Results }\label{sec:IV}


The CL$_{\rm s}$ technique \cite{cls1,cls2}
is exploited to find  upper limits on the signal cross section at $95\%$ CL.
Then the limits on the signal rates are translated into the upper limits on the LFV couplings.
In the CL$_{\rm s}$ method, we define log-likelihood functions  $L_{\rm Bkg}$ and $L_{\rm Signal+Bkg}$ 
for the background hypothesis, and for the signal+background hypothesis as the multiplication of
Poissonian likelihood functions. 
The $p$-value for hypothesis of signal+background 
and for the background hypothesis are determined using the log-likelihood
ratio $\rm Q =  - 2ln(L_{\rm Signal+Bkg}/L_{\rm Bkg} )$. 
The signal cross section is constrained using CL${_{\rm s} = \rm P_{Signal+Bkg}(Q > Q_{0})/(1-P_{Bkg}(Q < Q_{0})) \leq 0.05}$
 which is corresponding to $95\%$ confidence level  where $\rm Q_{0}$ is the expected value of test statistics $\rm Q$. 
 The {\tt RooStats} package \cite{roostats} is used to perform the numerical evaluation of the CL$_{\rm s}$.

\begin{figure}[htb]
\vspace{0.50cm}	
\begin{center}
{\includegraphics[ scale=0.35]{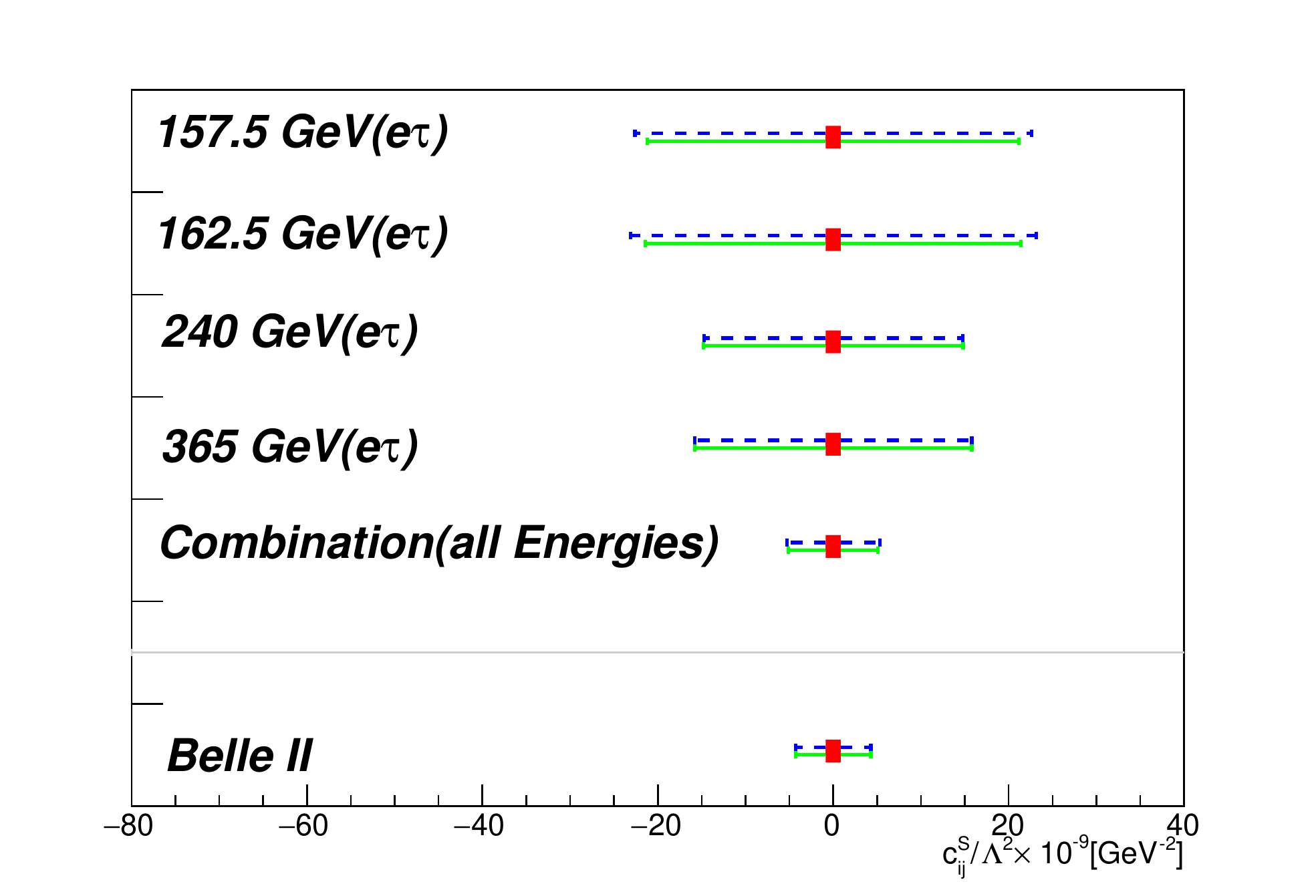}}
{\includegraphics[ scale=0.355]{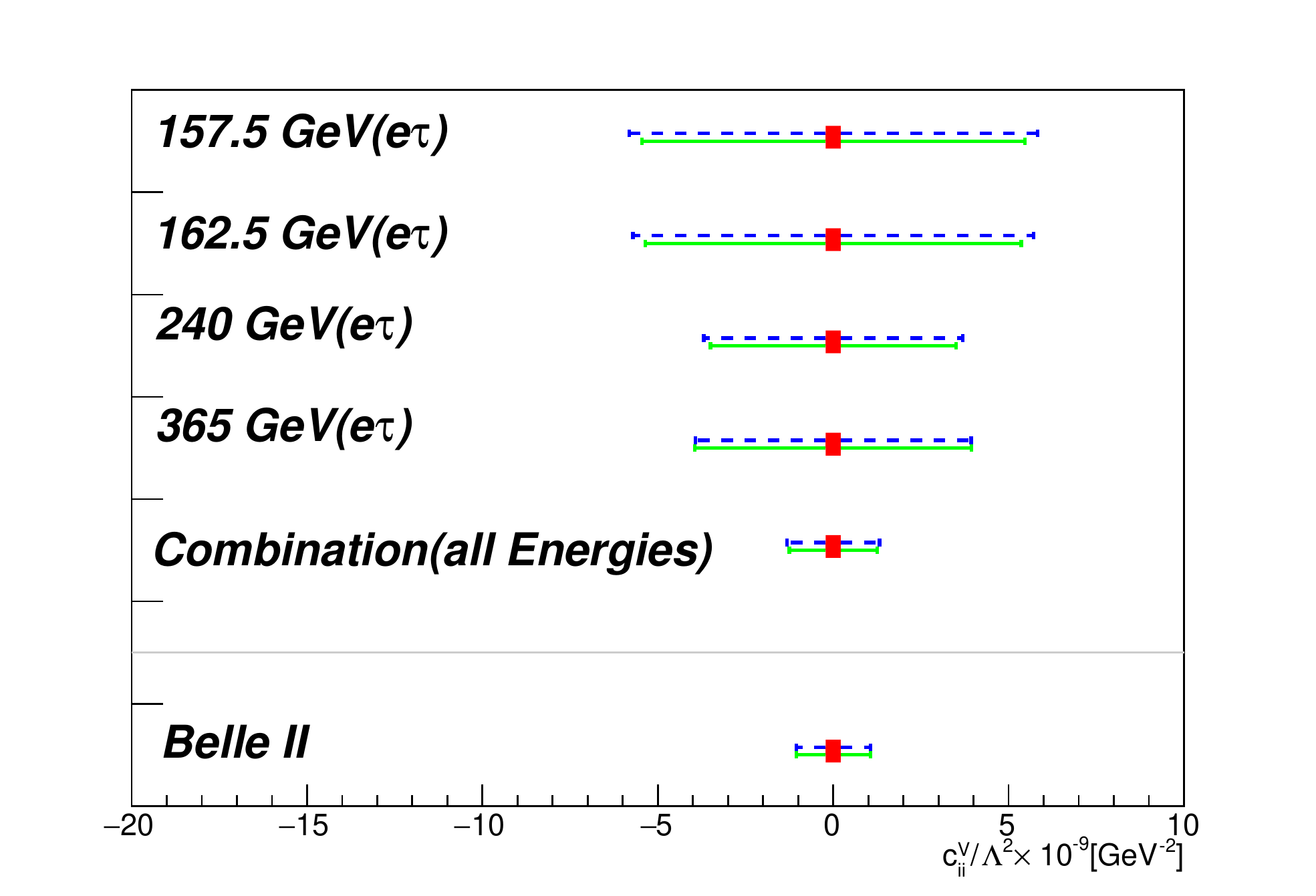}}
{\includegraphics[ scale=0.35]{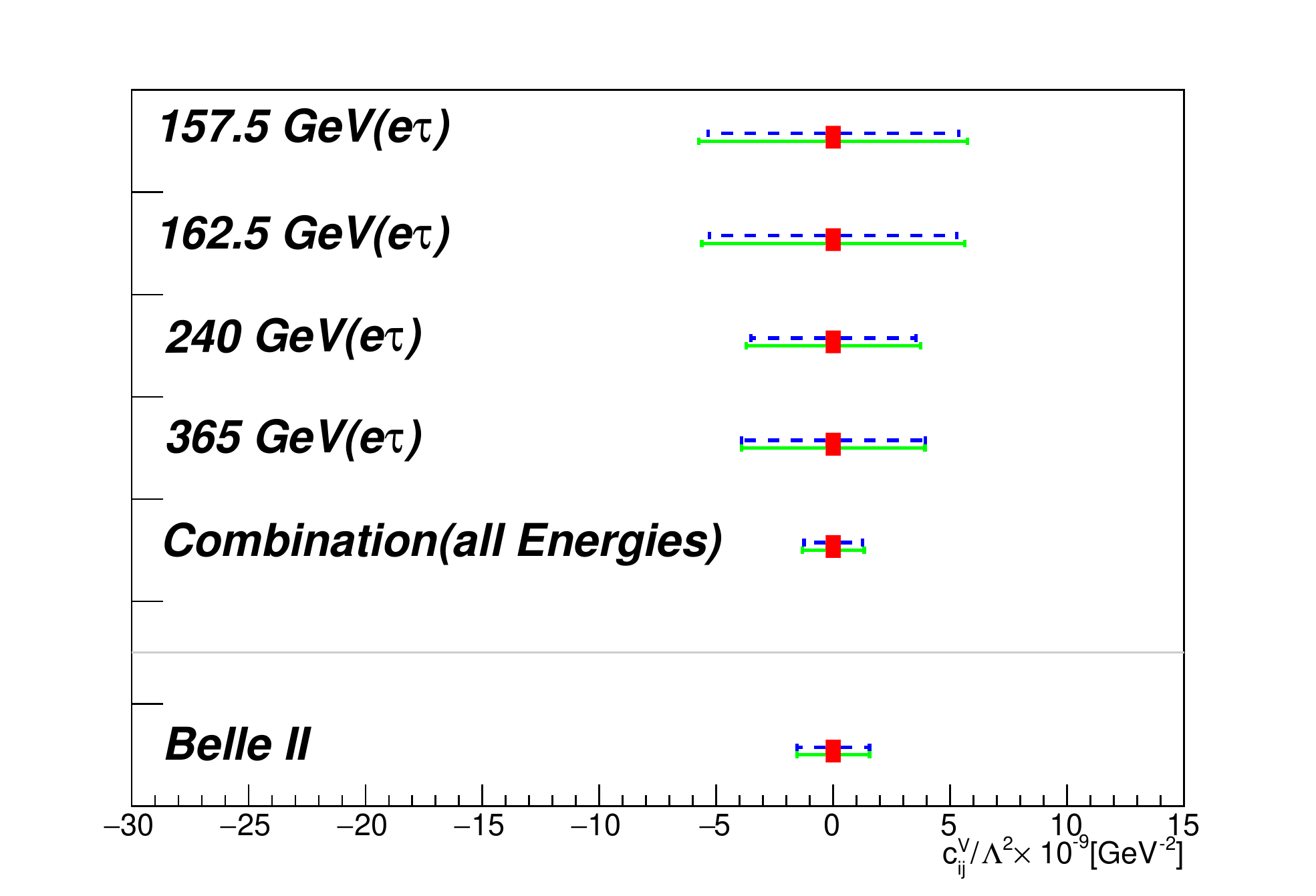}}
\caption{{\small Expected constraints at $95\%$CL  on $c^S_{ij}$ (top left) 
with  $c^S_{LR}$(dashed-blue) and $c^S_{RL}$(solid-green), $c^V_{ii}$ (top right) 
with $c^V_{LL}$(dashed-blue), $c^V_{RR}$ (solid-green), and $c^V_{ij}$ (bottom)  
with $c^V_{LR}$(dashed-blue) and $c^V_{RL}$ (solid-green)  from $e^-e^+\rightarrow e\tau$  
for four center-mass-energies are shown.  The result of combination of four energies and the prospects from
Belle II with $50\inab$ at $90\%$ CL \cite{r8} are presented for comparison.
} \label{graph:compare}}
\end{center}
\end{figure}

The predicted constraints at $95\%$ CL for $\sqrt{s} = 157.5 \GeV$ with 
$\mathcal{L}=5 \inab$, $\sqrt{s} = 162.5 \GeV$ with $\mathcal{L}=5 \inab$,  $\sqrt{s}$ = 240 GeV 
with $\mathcal{L}=5 \inab$ and $\sqrt{s} = 365 \GeV$ with $\mathcal{L}=1.5 \inab$  are presented in Table \ref{climits}.
These results are obtained considering only statistical uncertainties and the impacts of systematic and theoretical 
sources are neglected.  For illustration, the results are also given in Figure \ref{graph:compare}.
 As seen, the most sensitivity is achieved on vector type LFV couplings, i.e. on $c^V_{LL}$, 
 $c^V_{RR}$, $c^V_{RL}$, and $c^V_{LR}$. This is expected as the production rates for the
 vector type LFV couplings are larger than the scalar type by a factor of 16. 
 Among various center-of-mass energy scenarios, better sensitivity is expected to be obtained
 from $\sqrt{s} = 365$ GeV for which the signal cross section is largest as $\sigma(e^-e^+ \rightarrow e\tau) \propto s$. 
 Although, less amount of data is planned to be collected at $\sqrt{s} = 365$ GeV with respect to 
 the other energies, similar sensitivity to other energies at $\sqrt{s} = 365$ GeV is obtained.
 
To achieve better sensitivity, the results from four energy benchmarks are combined 
using the method explained in Ref. \cite{Khachatryan:2016vau}. The combined limits are
given in Table \ref{climits} and Figure \ref{graph:compare}. One can see that
the statistical combination of four energy scenarios improves the bounds by almost a factor of 
around 3 to 4 with respect to the results from a single energy benchmark.
It is informative to compare the results from prospects of Belle II with 
an integrated luminosity of 50 ab$^{-1}$. The limits from combination are competitive with 
those expected from Belle II.  Finally, we compare the results with those which are
expected from a future lepton collider at $\sqrt{s} = 1$ TeV with polarized beams
such that $P(e^{-}) = 0.8$ and $P(e^{-}) = -0.3$ \cite{lfv2}.  The results of combination are sensibly better for both the scalar and vector types of the LFV couplings.

\begin{table}[htbp]
	\begin{center}
		\scalebox{0.8}{
			\begin{tabular}{|c|c|c|c|c|c|c|}
				\hline	\hline				
				\tiny $\sqrt{s}$ (GeV) , $\mathcal{L}$ (ab$^{-1}$) & \tiny $\frac{c^V_{LL}}{\Lambda^2}[\times10^{-9}]$& \tiny $\frac{c^V_{RR}}{\Lambda^2}[\times10^{-9}]$   & \tiny $\frac{c^V_{RL}}{\Lambda^2}[\times10^{-9}]$ & \tiny $\frac{c^V_{LR}}{\Lambda^2}[\times10^{-9}]$ & \tiny $\frac{c^S_{RL}}{\Lambda^2}[\times10^{-9}]$ & \tiny $\frac{c^S_{LR}}{\Lambda^2}[\times10^{-9}]$ \\ 				
				\hline  \hline
				\small $157.5$  , $5$ 		 &\small$5.82$\small& \small$5.46$	&\small$5.74$ &\small$5.36$ &\small 21.18 &\small 22.61	\\ 
				\small $162.5$  , $5$		 &\small$5.71$& $5.36$	&$5.62$ &	$5.29$ &21.42 &23.12	\\ 	
				\small $240$    , $5$	        &$3.69$& $3.50$	&$3.73$ &	$3.53$ &14.81 &14.74	\\  									
				\small $365$   , $1.5$	        & 3.93 & 3.94 & 3.92 & 3.93 & 15.80 & 15.80	\\  \hline
				\small Combination             & 1.32 & 1.25 & 1.32 & 1.25 & 5.1 & 5.3	\\ \hline 
				\small Belle II		 	       & 1.06 & 1.06 & 1.55 & 1.55 & 4.29 & 4.29	\\ \hline	
				\small $\sqrt{s} = 1$	 TeV, pol. beam  &  4.3   &  1.1 &  1.6 & 1.8    &  13 & 5.9 \\ \hline    \hline	
			\end{tabular}}
			\caption{ \small The $95\%$ CL expected upper bounds on the scalar and vector type LFV couplings assuming four center-of-mass energies  
				as well as the combination are shown.  The limits are in the unit of $\GeV^{-2}$.
				The Belle II future prospects \cite{r8} and the expectation from  $e^{-}e^{+}$ 
				collider at $\sqrt{s} = 1$ TeV with polarized beam \cite{lfv2} are presented as well.
				\label{climits}} 
		\end{center}
	\end{table}

In order to have a feeling about the impact of systematic uncertainties on the results, we consider conservative 
values of uncertainties and re-estimate the sensitivities to the LFV couplings.
In Ref. \cite{opalz}, a search for LFV events at LEP2 with the OPAL detector using a similar final state
as this work has been performed. The analysis has used  the full  data  collected with OPAL 
at $\sqrt{s}$ between 189 GeV and 209 GeV. The systematic uncertainty on the signal efficiency is $3.5\%$
and on the number of expected background events is $5\%$. While for the future experiments, the measurements
are expected to be made with more accuracy, we consider a conservative value of $5\%$ uncertainty on both signal selection efficiency and on
background expectation. The  constraints  on $ c^V_{LL}/\Lambda^2$, $ c^V_{RR}/\Lambda^2$, $ c^V_{RL}/\Lambda^2$,
$ c^V_{LR}/\Lambda^2$, $ c^S_{RL}/\Lambda^2$, $ c^S_{LR}/\Lambda^2$, 
 derived at $\sqrt{s} = 365$ GeV, are found to be 
 $ 4.14 \times 10^{-9}$,  $ 4.15 \times 10^{-9}$, $ 4.12 \times 10^{-9}$, $ 4.13 \times 10^{-9}$, $ 16.57 \times 10^{-9}$, $ 16.65 \times 10^{-9}$ $\GeV^{-2}$. 
Comparing with the limits without systematic uncertainties shows that
including an overall $5\%$ uncertainty would not weaken the sensitivity remarkably.

\section{ Summary and conclusions}\label{sec:V}

Lepton flavor violation processes are absent in the SM but appear
in many extensions of the SM. In particular, the branching fractions of 
LFV $\tau^{\pm} \rightarrow \ell^{\pm} \ell^{\pm} \ell'^{\mp}$, $\ell, \ell' = e,\mu$
decays are increased in various beyond the SM scenarios.
In this work, the sensitivity of the future circular electron-positron collider, FCC-ee,
to probe the LFV couplings is examined using the $e^-e^+ \rightarrow e^{\pm}\tau^{\mp}$ 
production. To perform the study,  an effective Lagrangian approach, in particular, four fermi contact interactions with vector and 
scalar types are exploited. 
In order to find the sensitivity, events are generated for four run scenarios of center-of-mass energies of
$157.5$, $162.5$, $240$, and $365 \GeV$ with their corresponding benchmarks
for the integrated luminosity. 
The events are generated using {\tt MadGraph5\_aMC@NLO} considering ISR effect and passed through {\tt PYTHIA 8}
for showering, hadronization, and decay of unstable particles.  
A fast detector simulation is carried out by {\tt Delphes} using the ILD detector card.
The signal final state consists of an isolated electron and a $\tau$ lepton. In this study, 
the hadronic decays of the $\tau$ lepton are considered for which the branching fraction is around $64\%$. 
Based on the final state, the main sources of reducible and irreducible background processes are taken into account. 
Cuts on the energy of the final electron and the invariant mass of $e\tau$ system
are applied to suppress the background contributions. Upper limits at $95\%$ CL on the couplings
of various types of LFV couplings have been obtained using the CL$_{\rm s}$ method for the four center-of-mass energies.
 Finally, a statistical combination of results obtained at four center-of-mass energies is performed.
We show that the statistical combination of four center-of-mass energies  increases 
the sensitivity to the LFV couplings by a factor of two and larger with respect to the limits obtained at $\sqrt{s} = 365$ GeV. 
The results are compared with those obtained from other studies at lepton colliders considering beam 
polarizations and with prospects from Belle II. The results of combination are competitive with expectations for 
Belle II with 50 ab$^{-1}$.


\section*{Acknowledgments}
S.Tizchang is grateful to  Qiang Li for replying the question about implementing ISR in ${\tt MadGraph 5 \_aMC@NLO}$. S.M. Etesami is grateful to INSF for the financial support.	
R. Jafari and S. Tizchang are thankful to the Iran Science Elites Federation
for the financial support. 
	

\end{document}